\documentclass[12pt]{article}
\usepackage{amsmath,amssymb}
\batchmode

\newtheorem{lem}{Lemma}

\newcommand{\nn}{\nonumber}

\newcommand{\pa}{\partial}

\newcommand{\al}{\alpha}
\newcommand{\De}{\Delta}
\newcommand{\de}{\delta}

\newcommand{\Ga}{\Gamma}

\newcommand{\La}{\Lambda}

\newcommand{\Om}{\Omega}

\newcommand{\si}{\sigma}

\newcommand{\cC}{\mathcal{ C}}

\newcommand{\cH}{\mathcal{ H}}

\newcommand{\cF}{\mathcal{ F}}
\newcommand{\cK}{\mathcal{ K}}

\newcommand{\cL}{\mathcal{ L}}

\newcommand{\bbR}{{\mathbb{R}}}

\newcommand{\bbC}{{\mathbb{C}}}

\begin{document}

\title{Local String Field Theory}
\author{ 
J. Dimock\thanks{Research supported by NSF Grant PHY0070905}
\thanks{Talk delivered at ICMP 2003, Lisbon}\\
Dept. of Mathematics \\
SUNY at Buffalo \\
Buffalo, NY 14260 }
\date{}

\maketitle

\begin{abstract}
 We consider  open bosonic strings.  The non-interacting 
multi-string theory is described by certain free string field operators which we
construct.  These are shown to have local commutators with respect to a center of
mass coordinate.   The construction is carried out both in the light cone gauge and in 
a covariant formulation.
\end{abstract}

\newpage

\section{Overview}

A classical free    open  bosonic     string  in $\bbR^d$ is specified 
by a world sheet $X: \bbR  \times [0, \pi] \to  \bbR^d$
which satisfies the wave equation
\begin{equation}   \label{field}
( \frac {\pa^2}{\pa \tau^2 }- \frac {\pa^2}{\pa \sigma^2 })X^{\mu} ( \tau, 
\si)   =0 
\end{equation}
with  Neumann boundary conditions
\begin{equation}   \label{bc}
 \frac { \pa X^{\mu}}{ \pa \sigma} (\tau, 0) =
 \frac { \pa X^{\mu}}{ \pa \sigma} (\tau, \pi) =0  
\end{equation}
and  the constraint 
\begin{equation}  \label{constraint}
 \eta_{\mu\nu}( \frac { \pa X^{\mu}}{\pa \tau } \pm \frac { \pa X^{\mu}}{\pa \sigma })
( \frac { \pa X^{\nu}}{\pa \tau } \pm \frac { \pa X^{\nu}}{\pa \sigma }) = 0 
\end{equation}
where  $\eta_{\mu\nu} $ is the Minkowski metric in  $\bbR^d$ with  $\eta_{00} = -1,
\eta_{kk} = +1$ for  $k=1,\dots,d-1$.

For a  single  quantized string  in the Heisenberg picture one seeks an operator $X$
satisfying canonical commutation relations as well as the field equation and the constraint.
There are two standard ways to proceed.  On  the one hand one can use the constraint to eliminate
extra  variables  and then  proceed with canonical quantization.   This  only
works well when carried out in light cone coordinates and is known as the 
light cone gauge.   On the other hand one  can quantize directly ignoring the constraint
and then impose the constraint by insisting that wave functions be annihilated
by  constraint operators.  This is covariant quantization.   We discuss 
both in more detail below.

In each case we will be able to find a field equation satisfied by the 
wave functions  $\Psi = \Psi(X)$.  If  string coordinate  $X$ is  split  into a center of 
mass coordinate $x^{\mu}$ and internal coordinates  $ X^{'\mu} = X^{\mu} - x^{\mu}$, the 
wave function can be regarded as a function $\Psi  = \Psi(x,  X') $.
The field equation has the  form  of a Klein-Gordon equation  
\begin{equation}  \label{KG}
(-\square  + M^2)  \Psi =0
\end{equation}
where    $\square =  \eta^{\mu\nu} (\pa/\pa  x^{\mu}) (\pa/\pa  x^{\nu})$ is the d'Alembertian 
in the center of mass variable and 
$M^2$ is  a mass operator which acts on the internal variables  $X'$.

The string field theory describing many strings is obtained by introducing quantized field 
operators   $\Phi= \Phi(x,  X')$  obeying   $(-\square  + M^2)  \Phi =0$. 
This is second quantization.
We  carry out this construction  in both the light cone gauge and the covariant theory.
Our main interest is to show that the fields have a vanishing commutator when the 
center of mass coordinates are spacelike separated.  Formally this is 
\begin{equation}
 [\Phi(x,  X'),  \Phi(y,Y')]
=0   \ \ \ \ \   \textrm{if     }  (x-y)^2 >0
\end{equation}
and we will give a precise version.   This report summarizes the results
of two papers  \cite{Dim00}, \cite{Dim02}. Earlier results in the physics 
literature can be found in 
 \cite{Mar93}, \cite{Low94}, \cite{LSU94},  \cite{HaOd97}.
\bigskip

\noindent
\textbf{Notation: }  One can write the two-dimensional wave equation
 as a  first order system
\begin{equation}      \label{wave}
\frac{\pa  X^{\mu}}{\pa \tau}  =  P^{\mu}  \ \ \ \ \ \ \ \ \
\frac{\pa  P^{\mu}}{\pa \tau}  =  \frac{\pa ^2 X^{\mu}}{\pa \si^2} 
\end{equation}
Suppose we expand  in a cosine series as dictated by the boundary conditions.
The coefficients are the center of mass coordinates
\begin{equation}
x^{\mu} (\tau) =    \frac{1}{\pi} \int_0^{ \pi} X^{\mu} (\tau, \sigma ) d \si 
  \ \ \ \ \ \ \ \ \
p^{\mu} (\tau) =   \frac{1}{\pi} \int_0^{ \pi} P^{\mu} (\tau, \sigma ) d \si 
\end{equation}
and the internal coordinates for  $n=1,2, \dots$
\begin{equation}
  x_n^{\mu}(\tau) = \frac{\sqrt{2}}{\pi} \int_0^{ \pi} X^{\mu} (\tau, \sigma )
\cos n \si d \sigma   \ \ \ \ \ \ \ \ 
p_n^{\mu}(\tau) = \frac{\sqrt{2}}{\pi} \int_0^{ \pi} P^{\mu} (\tau, \sigma )
\cos n \si d \sigma   
\end{equation}
These satisfy
\begin{equation}
\begin{split}
dx^{\mu} / d \tau = p^{\mu}  &  \ \ \ \ \ \ \ \ \  dp^{\mu} / d \tau = 0 \\
dx^{\mu}_n / d \tau = p^{\mu}_n  &  \ \ \ \ \ \ \ \ \  dp^{\mu}_n / d \tau = -n^2 x^{\mu}_n \\
\end{split}
\end{equation}

\section{Light cone gauge}

Assuming $X$ satisfies the wave equation,  boundary conditions, and constraints,
  we still have many  possibilities for parameterizing the solution.  To take 
advantage of this we 
change to 
light cone coordinates defined by mapping  $x = ( x^0,\dots,
x^{d-1})$ to  $(x^+,x^-, \tilde x)$ where
\begin{equation}
x^{\pm}  = ( x^0 \pm x^{d-1})/ \sqrt{2}\ \ \ \ \ \ \ \  \tilde x  =  ( x^1,\dots, x^{d-2})
\end{equation}
A solution 
is said to be  in the \textit{light cone gauge} if   $X^+(\tau,\sigma) = p^+ \tau$.  As to
the existence of this gauge we have the following:

\begin{lem}\nonumber
Let  $ X^{\mu}(\tau, \sigma) $ satisfy
the wave equation,  boundary conditions,  and constraints and 
\begin{equation} \label{plus}
P^+ = \frac{\pa X^+}{\pa \tau} >0 
\end{equation}
Then there exists a conformal diffeomorphism on $(\tau,\sigma)$
such that in the new coordinates  all these conditions  still 
hold and in addition $X^+ = p^+ \tau$  with  $p^+ >0$
\end{lem}

The condition  (\ref{plus}) says that $\tau$ is a  forward moving parameter
in a certain sense. 
For the proof see  \cite{Dim00}. 

Suppose then  we have a solution in the light cone gauge.
This  choice of gauge  and the constraints can be  used to 
eliminate      $x^+,p^-,x_n^{\pm}, p_n^{\pm}$ as dynamical variables.
  If we take  $x^+  =  p^+
\tau$  as the time parameter we find that  the dynamical equations have become
\begin{eqnarray}
 d x^{-}/ d x^+  = p^{-}/p^+ &\ \ \ \ & d p^+ /d x^+   = 0 \nn \\
  d x^{k}/ d x^+  = p^{k}/p^+ &\ \ \ \ & d p^k /d x^+   = 0 \nn \\
 d x^{k}_{n}/d x^+   = p_{n}^k/p^+   & \ \ \ \  &
d p_{n}^{k} /d x^+   =   (-n^2/p^+)  x^k_{n}   
   \label{xeqn} 
\end{eqnarray} 
This is a Hamiltonian system with   Hamiltonian 
\begin{equation}
 p^- = 
  \frac {1} { 2 p^+ }\left( \tilde p^2 +
\sum_{n=1}^{\infty} (\tilde p_n^2  + n^2 \tilde x_n^2 )\right)
\label{pm}
\end{equation}

\bigskip

Now we quantize this system by imposing canonical commutation relations 
on the variables $(p^+,x^-), (x^k,p^k), (x^k_n,p^k_n)$. 
These can be realized as operators on a Hilbert  space of Fock valued
functions 
\begin{equation}
\cL^2(\bbR^+ \times   \bbR^{d-2} , \cF,  d p^+ d \tilde p / 2p^+)
\end{equation} 
The operators  $p^+,  p^k$ are multiplication operators.
The operators $ x^k_{n}  , p^k_{n}$  
act on the  Fock space of  
transverse modes
\begin{equation}
\cF  = \cF ( \cL^{2, \perp}([0,\pi],\bbC^{d-2}))
\end{equation}
Here $ \cL^{2, \perp}$ means the subspace of $ \cL^{2}$ orthogonal to the constants and 
\begin{equation}
 x^k_{n}  =  (2n)^{-1/2} ((a^k_{n})^*+a^k_{n}) \ \ \ \ \ \ \ \ 
p^k_{n}   =  i (n/2)^{1/2} ((a^k_{n})^*-a^k_{n})
\end{equation}
where   $(a^k_{n})^*$ is the creation operator for 
 function $(0,\dots,\sqrt{\frac{2}{\pi}}\cos n \sigma,\dots,0)$ (entry in the $k^{th}$ slot).

The quantum  Hamiltonian   $p^{-}$  then  has the form 
\begin{equation}
 p^- = 
  \frac {1} { 2 p^+ }\left( \tilde p^2 +  M^2\right)
\end{equation}
where 
\begin{equation}
\begin{split}
M^2 \equiv & \sum_{n=1}^{\infty} (:\tilde p_n^2:  + n^2
:\tilde x_n^2 :)  -2a  \\
&=  2  \left(\sum_{n=1}^{\infty}  \sum_{k=1}^{d-2} n (a^k_n )^*(a^k_n )-a\right)\\
&=   2  \left(\sum_{n=1}^{\infty}  \sum_{k=1}^{d-2} \al_{-n}^k \al_n^k -a\right)\\  
\end{split}
\end{equation}
Here we have Wick ordered  and allowed an adjustment by an
arbitrary constant  $2a$.  In the last line we have introduced  the
notation more common in string theory  $\al^k_n = -i \sqrt{n} a^k_n$
and  $\al^k_{-n} = i \sqrt{n} (a^k_n)^*$.
The operator    $M^2$  on  $\cF$ is a sum of harmonic 
oscillators. It is self-adjoint and 
has  spectrum   $-2a,\ 2-2a\ ,4-2a\ , \dots$.
The operator  $p^-$ is also self adjoint. 

In  the Schrodinger picture  our Fock-valued 
 wave functions evolve according to 
\begin{equation}
\Psi(x^+, p^+,  \tilde p)
=   e^{-ip^-x^+}   \Psi  ( p^+,  \tilde p)
\end{equation}
In configuration space this becomes 
\begin{equation}
\Psi(x^+,x^-, \tilde x)  =  (2\pi)^{-(d-1)/2} \int e^{-ip^-x^+ -ip^+x^-+i \tilde p \tilde x}
\Psi( p^+,  \tilde p) d p^+ d \tilde p /2p^+
\end{equation}
  which satisfies  the Klein-Gordon equation
\begin{equation}
(2 \pa_+ \pa_{-}  -  \tilde \De  + M^2 )  \Psi  =0
\end{equation}
This shows that  $M^2$ can be interpreted as a mass operator for the string.
This will be our field equation.

One can now ask  whether the theory is  Lorentz covariant. This 
is difficult because of the special choices that have been made.  Nevertheless it
is formally true provided $d=26$ and $a=1$.  We do not  pursue
this, but instead  turn to the covariant
theory where Lorentz covariance comes naturally.

\section{The covariant theory}

In the covariant theory we seek a quantization  without making
special choices of coordinates.  Now  $X^{\mu}$ and 
  $P^{\mu}=\pa  X^{\mu}/\pa \tau  $ are quantized  by solving  
the wave equation with  the commutation relations at  $\tau =0$:
\begin{equation}
[X^{\mu}(\si ), P^{\nu}(\si')] = i \pi\de (\si - \si') \eta^{\mu \nu}  
\end{equation}
We jump right to the solution which is  
\begin{equation}  \label{X}
 X^{\mu}(\tau,\sigma) = x^{\mu} + p^{\mu} \tau   + i\sum_{n \neq 0}\al^{\mu}_n
 e^{-in \tau} \frac{\cos n \sigma }{ n}    
\end{equation}
where the center of mass operators   $x^{\mu}, p^{\mu}$ 
and the internal  operators 
$\al_n^{\mu}$ are required to 
satisfy the
commutation relations
\begin{equation}
[x^{\mu}, p^{\nu}] =   i \eta^{\mu \nu}  \ \ \ \ \ \ \ \ \
 [\al_m^{\mu}, \al_n^{\nu}] = m \de_{m+n}  \eta^{\mu \nu}\label{CCR1}  
\end{equation} 
Operators satisfying these relations can  be constructed on
a Hilbert space of the form  
\begin{equation}
\cL^2 ( \bbR^d ,  \cF, dp )  
\end{equation}
with  $p^{\mu}$ as
a multiplication operator and the $\al_n^{\mu}$ as creation and
annihilation operators   on the Fock space  
\begin{equation}
\cF = \cF\left(\cL^{2, \perp} (  [0,\pi], \bbC^d )\right)
\end{equation}
These spaces have indefinite inner products, for example
on  $\cL^2 (  [0,\pi], \bbC^d )$  the inner product is 
\begin{equation}
  <f,g>  = \int_0^{\pi} \eta_{\mu \nu}\overline{ f^{\mu}(\si)} g^{\nu}(\si) d \si
\end{equation}

Now consider the constraint equations which we want to impose as
a condition on the wave functions.      
Passing to  Fourier components  and Wick ordering  one finds that   
the conditions are    $L_n  \Psi =0$  for integer $n$  where: 
\begin{equation}  \label{L}
\begin{split}
 L_0 &= \frac12 \  p^2 + \sum_{n=1}^{\infty}\al_{-n} \cdot \al_n
 \\
 L_m &= \al_m \cdot p \  + \ \frac12 \sum_{n \neq m,0}\al_{m-n} \cdot \al_n \ \ \ \ \ \  m \neq 0
  \\
\end{split}
\end{equation}
We allow a shift in $L_0 \to  L_0-a$, put aside the 
constraints for  $m<0$   (a standard compromise), and ask that  
\begin{equation}
(L_0 - a) \Psi = 0  \ \ \ \ \ \ \ \ \ \ 
L_m \Psi = 0 \ \ \ \ \ \ \ \ \ \ \ m>0  
\end{equation}

These constraints  cannot  be imposed naively  since  $p^2$ has continuous spectrum and 
$ \sum_n\al_{-n} \cdot \al_n $ has discrete spectrum. 
To work around this 
we first  decompose   our Hilbert space as a direct integral
\begin{equation}  \label{1}
\cL^2(\bbR^d, \cF, dp ) = 
 \int ^{\oplus} \cL^2(V_r, \cF, d \mu_r)\ dr
\end{equation}
where   $V_r = \{p:  p^2+r =0\}$ is the mass shell and $\mu_r$ is the Lorentz 
invariant measure on $V_r$.  Then  $L_0$ and  $L_m$ are decomposable 
and we have $L_0-a =  \int ^{\oplus}\frac12 (-r + M^2)\ dr$  where 
$M^2$ on  $\cF$ is given by 
\begin{equation}
M^2= 2 \sum_{n=1}^{\infty}\al_{-n} \cdot \al_n -2a
\end{equation}
This  again has spectrum  $ -2a, 2-2a, \dots$.
To get a nontrivial null space for  $L_0-a$  we pick  the 
values $r= -2a, 2-2a, \dots$
out of the  
direct integral   and form the direct sum  
\begin{equation}
\cH  =  \bigoplus_{r} \cL^2(   V_r^{(+)}, \cF,  d\mu_r)    \equiv  
\bigoplus_{r} \ \  \cH_r
\end{equation}
Then  $L_0, L_m$ act on this space. On a 
vector   $\Psi =   \{ \Psi_r \}$  the constraints  
 are
\begin{equation}
M^2\Psi_r = r \Psi_r  \ \ \ \ \ \ \ \ \ \ \  L_m \Psi_r =0
\end{equation} 
which can be satisfied.

Let  
\begin{equation}
\cH'=\bigoplus_{r} \cH'_r  
\end{equation}
 be the subspace of  $\cH$ satisfying the constraints.
We divide out isotropic elements
\begin{equation} 
\cH'' = \cH'  \cap  (\cH')^{\perp}  =  \bigoplus_{r} \cH''_r  
\end{equation}   
which are null vectors, 
and form the physical Hilbert space 
\begin{equation}
\cH^{phys}  =  \cH' / \cH''  = \bigoplus_{r}  \cH^{phys}_r
\end{equation}
The famous ``no-ghost" theorem (see for example \cite{FGZ86}) asserts that 
the inner product is positive definite on  $\cH^{phys}$  provided
$d=26$ and $a=1$. We make this choice, so that the sum is over  $r=-2,0,2,\dots$
The prize  for all this is   that one now 
has   a natural  unitary representation  $U(a, \La)$ of the inhomogeneous Lorentz
group.

Let us exhibit some physical states.  If  $\Om_0$ is the empty state 
in $\cF$  and  $f  \in  \cL^2(V_{-2},\bbC, d\mu_{-2})$    then  
\begin{equation}
\Psi(p)  = f(p) \Om_0 
\end{equation}
is an element of  $\cH_{-2}$
 It satisfies  $M^2 \Psi =-2\Psi$ and $ L_m \Psi =0$
and so is an element of    in  $\cH'_{-2}=\cH^{phys}_{-2}$.  This is  a scalar of mass
$-2$  called the \textit{tachyon}.   If  $f  \in  \cL^2(V^+_{0},\bbC^d, d\mu_0)$ 
then 
\begin{equation}
\Psi(p)  = f_{\mu}(p)\al^{\mu}_1 \Om_0 
\end{equation}
is an element of  $\cH_{0}$.  It  satisfies   $M^2 \Psi  =0$ and  if
   $p^{\mu}f_{\mu}(p)=0$  it satisfies  $L_m\Psi=0$ as well and hence is 
an element of   $\cH'_0$.  Dividing by $\cH''_0$ removes longitudinal states with 
$f_{\mu}(p) = p_{\mu}h(p)$ and we get  elements of  $\cH^{phys}_0$.  These
are identified as photons and this is essentially  the Gupta-Bleuler construction.
 Higher mass physical states can also be exhibited, see \cite{Dim02}

Finally consider a wavefunction  $\Psi  =  \{ \Psi_r \}$ with at least the first
constraint  $M^2\Psi_r= r \Psi_r$ satisfied. 
Then  the Fourier transform 
\begin{equation}
 \Psi(x)   =  \sum_r  \int_{V_r}  e^{-ip \cdot x}  \Psi_r(p)  d\mu_r(p)
\end{equation}
again satisfies the Klein-Gordon equation
\begin{equation}
(-\square + M^2 )   \Psi  =0
\end{equation}

\newpage

\section{String fields}

Now we proceed to quantize the two  field equations we have identified.
Although we have arrived at them in quite different ways
they both have the form of a Klein-Gordon equation
 $(-\square  + M^2)  \Phi =0$ for  functions  $\Phi:\bbR^d  \to  \cF$.
(We revert to standard coordinates for the light
cone gauge).
The difference is in the Fock space  $\cF$.  We have 
\begin{equation}
\cF =  \left \{  \begin{array}{lcl}
\cF\left( \cL^{2, \perp}([0,\pi],\bbC^{d-2})\right)& & \textrm{light cone gauge}\\
\cF\left( \cL^{2, \perp}([0,\pi],\bbC^{d})\right)& & \textrm{covariant theory}
\end{array}  \right.
\end{equation}
In the second case there is an indefinite inner product
and the remaining constraints  $L_m \Phi =0$ must still  be satisfied,
The mass operators $M^2$  on $\cF$ are   
different but have the same spectrum.
We discuss the two cases in parallel.

First we consider the ``classical" equation, i.e.  $(-\square  + M^2) U =0$ for functions 
$U$ which are real valued with respect to some conjugation on  $\cF$.
These equations have   advanced and
retarded  fundamental solutions 
$E^{\pm}$ which are defined on   test functions    $F \in  \cC^{\infty}_0(  \bbR^d,
\cF) $. They satisfy  $(- \square + M^2)E^{\pm}F = F$ and $E^{\pm}F$
has support in the causal future/past of the support of   $F$.  Explicitly they are  given
by 
\begin{equation}
(E^{\pm} F)(x)  = \frac{1}{(2 \pi)^{d/2}}  \int_{\Ga_{\pm} \times \bbR^{d-1}}
 \frac{ e^{ip\cdot x}}{p^2 + M^2} \ 
\tilde  F(p) dp 
\end{equation}
where the  $p^0$ contour $\Gamma_{\pm} $ is the real line shifted slightly above/below
the  real axis.  We will also need the propagator function  
\begin{equation}
E = E^+ - E^-
\end{equation}
Then $U =EF$ solves   $(-\square  + M^2)  U =0$ and  has $\cC^{\infty}_0$  Cauchy data 
on any spacelike hypersurface.  We say it is a  \textit{regular} solution.
Conversely any regular solution $U$ has the form  $U= EF$.

 If  $U,V$ are regular solutions then 
\begin{equation}
\sigma (U,V)  =  \int_{x_0 = t }( <U(x), \frac{\pa  V}{ \pa x^0}(x) >-
  < \frac{\pa U}{ \pa x^0}(x)  ,V(x)>)\ d \vec x
\end{equation}
is independent of $t$  by Green's identity. For  definiteness
take    $t=0$.  Also  by Green's identity
any  regular solution
$U$ regarded as a distribution can be expressed in terms of its Cauchy 
data by  
\begin{equation}   \label{evolve}
<U,F>  = \sigma(U,EF)
\end{equation}

Now we turn to the quantized version.
The phase space we want to quantize is the space of  Cauchy data for $(-\square  +
M^2)  U =0$ on a spacelike hypersurface,  or equivalently  the    space of  regular 
solutions $U$ itself.  The form $\sigma (U,V) $ is the natural symplectic 
form on the space and quantization consists of finding operators   $\sigma(\Phi,U)$
indexed by  regular solutions  $U:  \bbR^d  \rightarrow  \cF$  such that 
\begin{equation}
[\sigma(\Phi,U),\sigma(\Phi,V)]  = -i\sigma(U,V)
\end{equation}
These are the canonical commutation relations  (CCR). 
Given a representation of the CCR we create
a string field operator  with test functions   $F \in  \cC^{\infty}_0(  \bbR^d,
\cF) $ as  in  (\ref{evolve}) by 
\begin{equation}
\Phi(F)  = \sigma(\Phi, EF)  
\end{equation}
Then   $\Phi$ satisfies the field equation in the sense of distributions,
\begin{equation}  \label{one}
\Phi( (-\square  +M^2)  F )=0
\end{equation}
and  using $\sigma(EF,EG) =  -<F,EG>$ it has the commutator  
\begin{equation}  \label{two}
[\Phi(F), \Phi(G)] =  -i <F,EG>  
\end{equation}
This is the structure we want, and in fact one can show that any 
 operator valued distribution 
$\Phi(F)$ satisfying (\ref{one}),(\ref{two}) arise from a representation
$\sigma(\Phi,F)$  of the CCR in this manner.
The  locality result is  now immediate.   If  $F,G  \in   \cC^{\infty}_0( 
\bbR^d,\cF)$ have spacelike separated supports then $<F,EG>=0 $
and hence
\begin{equation}
[\Phi(F), \Phi(G)] =  0  \ \ \ \  \textrm{(locality)}
\end{equation}
This  completes the abstract discussion for the light cone gauge, but
for the covariant theory there are still constraints to be satisfied 
and the interpretation of the constraints seems to  depend on 
the representation.

What representations of the CCR should we consider? What
representations might have physical relevance? 
This  is not clear.

If we  suppress  tachyons there is a distinguished positive energy 
representation we can consider.   The \textit{ad hoc} suppression of tachyons  is not really
satisfactory, but it does give some insight  and makes contact with point field theory.
We give some details in the covariant case; the light cone gauge is
similar.
 Excluding $r=-2$ we   define   as before 
\begin{equation}
\cH_+ =  \bigoplus_{r=0,2,...} \cH_r  =   \bigoplus_{r=0,2,...}  \cL^2(V_r^+, \cF, d \mu_r)
\end{equation}
For    $F  \in   \cC^{\infty}_0( \bbR^d,\cF)$  define 
$\Pi_+  F  \in  \cH_+$ by   
\begin{equation}
(\Pi_+ F)_r(p)    =  \sqrt{ 2 \pi} \ P_r \tilde F ( p )   \ \ \ \ \   p \in V_r^+
\end{equation}
Here   $P_r$ 
is the projection onto the $M^2=r$ subspace of  $\cF$. 
Let   $a,a^*$ be creation and annihilation operators on the Fock space
$\cK  = \cF(\cH_+)$ and   define 
\begin{equation}
\Phi(F)  =  a( \Pi_+ F)  + a^* (\Pi_+ F)
\end{equation}
This satisfies   (\ref{one}),  (\ref{two}) and  generalizes
the positive energy representation for point fields.

Continuing with this positive energy representation we 
impose the constraint by asking for states annihilated by 
$L_m \Phi$, actually just  the annihilation part  of this operator.
This turns out to be  $\cK' =  \cF(\cH'_+)$  and dividing out 
isotropic elements gives    
$\cK^{phys} =  \cF(\cH_+^{phys})$  which has a positive definite  inner product. 
 The field
operators  $\Phi(F)$ act on $\cK^{phys}$ if   $\Pi_+ F \in \cH'_+$.  We call
such fields  \textit{observable fields}. 
The observable fields
still have the local commutator $[\Phi(F), \Phi(G)] =  -i <F,EG>  $.
However now it is not clear whether this can be made to vanish,
i.e. it is not clear whether one can satisfy  $\Pi F \in \cH'_+$
 and still have some control over the support of $F$.  Thus the existence 
of local observables is not settled in this covariant case;  in the light
cone gauge there is no problem.

\section{Comments}  

\begin{enumerate}
\item
There is an interesting extension of these results \cite{Dim00},
originally  due to Martinec  \cite{Mar93}.
Consider the light cone gauge and change from a Fock representation
for the internal degrees of freedom to a Schrodinger representation.
The  $x_n^k$  are now independent Gaussian random variables with 
mean zero and 
variance $(2n)^{-1}$ and     $p_n^k  =  -i \pa/ \pa x_n^k  + inx_n^k$.
The field equation $(- \square +M^2) U =0$ now takes  
form
\begin{equation}
\left( (\frac {\pa} { \pa x^0} )^ 2  -\sum_{k=1}^{d-1} 
 (\frac {\pa} { \pa x^k} )^ 2    
- \sum_n   \sum _{k= 1}^{d-2}  
\left((\frac {\pa} { \pa x^k_{n}} )^ 2  
 -2nx_n^k \frac {\pa} { \pa x^k_{n}}\right) -2 \right)  U = 0 
\end{equation}
Suppose we consider   functions  $U = U(x^{\mu}, \{  x^k_{n}\} )$  which depend only
on a finite number of these modes, and so restrict  the sum over  $n$  to  $n\leq N$.
Then we have a  strictly hyperbolic differential equation  with domain of 
dependence defined by the metric 
\begin{equation}
  - (dx^0) ^2 + \sum_{k=1}^{d -1} (dx^k)^2 +  \sum_{n = 1}^N \sum _{k= 1}^{d-2} 
(dx^k_{n})^2  
\end{equation}
Test functions for field operators, formerly Fock valued, can now be regarded  as functions
$F = F(x^{\mu}, \{  x^k_{n}\} )$ of these variables.  Then one can show that if
$F,G$ have spacelike separated supports with  respect to the above metric
then  \[[\Phi(F), \Phi(G)] = 0\]   One says that  the field is local with
respect to the \textit{string light cone}.  This is a limitation on how 
fast the various modes can grow.
\bigskip

\item  Interacting string field theory does not exist, although there
are candidates.  Is there any chance that such a theory also has 
a locality property?   For some speculation in
this direction  see  \cite{LSU94}

\end{enumerate}

\end{document}